# Metal-Insulator Transition and Anomalous Lattice Parameters Changes in Ru-doped VO$_2$

*Xin Gui and Robert J. Cava*[*]

Department of Chemistry, Princeton University, Princeton NJ 08540, USA

***ABSTRACT***

VO$_2$, of interest for decades due to both its phenomenology and its potential applications, has a monoclinic distortion of the rutile crystal structure at ambient temperature that is coupled to its metal-insulator transition (MIT). In contrast, RuO$_2$ has three electrons more per formula unit, is a metallic conductor, and has an undistorted rutile structure. Here we report a systematic study of Ru-doped VO$_2$ (V$_{1-x}$Ru$_x$O$_2$, $0.01 \leq x \leq 0.9$); generally characterizing its crystal structure, magnetic and electronic properties, and heat capacity. The composition-dependent Wilson ratio is determined. We find that an unusually high Ru doping value (80%, x=0.8) is required to achieve a metallic state in V$_{1-x}$Ru$_x$O$_2$. No superconductivity was observed down to 0.1 K in the metallic materials. We propose a possible understanding for how the insulating state can exist in V$_{1-x}$Ru$_x$O$_2$ at high Ru contents.

*Address correspondence to: rcava@princeton.edu

## 1. Introduction

Vanadium dioxide ($VO_2$) undergoes a metal-insulator transition (MIT) when cooled through ~340 K, accompanied by a structural phase transition from a high-temperature tetragonal structure to a low-temperature monoclinic structure.[1,2] It has been widely investigated for decades due to this MIT, which is of strong fundamental interest,[3–29] and due its potential in applications, which include optical switches,[30] strain sensors[31] and gas sensors[32]. The MIT in $VO_2$ has been investigated by heating,[4] doping,[21,22,24,25,33] application of electric fields[34–36] and structural stress[14,37,38], for example. Not without controversy, both the Peierls state, which involves spin singlet formation by electrons on neighboring V atoms, and the Mott-Hubbard state, for which electrons attempting to occupy the same site undergo coulombic repulsion, have been proposed as needed to explain the insulating state.[5,11,16–20,39–42]

Here we report a systematic study of how Ru substitution for V in $VO_2$ affects its crystal structure, magnetic and electronic properties and heat capacity. The lattice parameters of Ru-doped $VO_2$ have been previously reported, but without any characterization of its physical properties.[43] Since Ru has three more valence electrons than V, it can be imagined that the metallic state should be easily induced by doping a small amount of Ru into $VO_2$. However, our results show that the insulating phase in $VO_2$ does not become metallic until a high doping value of 80%, i.e., that $V_{0.2}Ru_{0.8}O_2$ is a metal while $V_{0.3}Ru_{0.7}O_2$ is not. We propose a pathway for how Ru doping can affect the magnetic and electronic properties in $V_{1-x}Ru_xO_2$.

## 2. Experiment

**2.1. Preparation of polycrystalline samples of $V_{1-x}Ru_xO_2$:** Polycrystalline $V_{1-x}Ru_xO_2$ (0 < x < 1) was synthesized by using high-temperature solid-state method. $VO_2$ (99%, powder, Beantown Chemical) and $RuO_2$ (99.9%, powder, Sigma-Aldrich) were used as starting materials as purchased. The mixtures of $VO_2$ and $RuO_2$ with appropriate stoichiometry were thoroughly ground and placed in alumina crucibles which were then sealed in evacuated quartz tubes. Two heat treatments at 950 ºC for two days were carried out, with intermediate grinding. The final products were black powders and are resistant to air and moisture.

**2.2. Phase Identification:** Powder X-ray diffraction (PXRD) characterization of the materials was carried out on a Bruker D8 Advance Eco diffractometer with Cu Kα radiation and a LynxEye-XE detector. The lattice parameters of $V_{1-x}Ru_xO_2$ were determined by using Rietveld fitting within the FullProf Suite.

**2.3. Physical Property Measurements:** Magnetization measurements were performed using a Physical Property Measurement System (Quantum Design PPMS) with a vibrating sample magnetometer (VSM). M vs T data were collected in an applied field of 3000 Oe in the temperature range of 1.8 K to 350 K, and, when necessary, under various applied magnetic fields. The magnetic susceptibility was defined as M/H where H is the applied magnetic field in Oe and M is the measured magnetization in emu. The four-probe method was employed to measure temperature-dependence of resistivity. The samples for resistivity measurements were prepared by pressing $V_{1-x}Ru_xO_2$ powders and annealing them at 1000 °C for 12 hours. The phase was not changed after this process, confirmed by PXRD. Heat capacity was measured using a standard relaxation method in the PPMS from 2 K to 30 K. Samples used for heat capacity measurements were pellets of the powders annealed at 950 °C.

## 3. Results and Discussion

**3.1. Crystal Structure and Phase Determination:** Undoped $VO_2$ above ~340 K crystallizes in the rutile structure. This structure has tetragonal symmetry, in space group $P\,4_2/mnm$ (No. 136)[9,10]. The undoped material undergoes a structural phase transition to monoclinic space group $C\,2/m$ (No. 12)[2] accompanied by a metal-insulator transition (MIT) upon cooling below ~340 K. With electron doping, in Mo-doped $VO_2$ for example,[21] the MIT is suppressed and the material adopts a tetragonal structure below 300 K. The tetragonal structure of $VO_2$ is shown in Figure 1(a). Although there is only one crystallographically equivalent site for either V or O in the tetragonal structure, to facilitate comparison to the monoclinic form, we employ different colors to represent both elements. Edge-shared $VO_6$ octahedra stack along the $c$ axis of the unit cell and form quasi-one-dimensional vanadium chains running parallel to $c$, as shown at the bottom of Figure 1(a). The $VO_6$ octahedra share oxygen atoms on their vertices perpendicular to the chains. The interatomic distances for V1-O2 and V2-O1 are the same while V1-O1 and V2-O2 have the same bond lengths and the latter are longer than the former.

Similar to the case for Mo-doped $VO_2$, heavily doped $V_{1-x}Ru_xO_2$ also stabilizes the tetragonal structure at room temperature. Powder XRD patterns of x = 0.01 and 0.02 are shown as

Figure S1 in the Supporting Information (SI).[44] They have been fitted by Le Bail method and are found to adopt a triclinic space group $P$ -1 (No. 2). Figure 1(b) presents the powder XRD patterns of $V_{1-x}Ru_xO_2$ at ambient temperature with x ranging from 0.03 to 0.90. The obtained patterns are in good agreement with the tetragonal rutile structure. (Results for all compositions are shown in Figure S2 in the SI.)[44] As shown in the insets of Figure 1(b), the peak positions do not shift monotonically with increasing amount of dopant. For instance, the (0 0 2) peak shifts slightly to higher angle when x ≤ 0.15 while it shifts slightly to lower angle when 0.2 ≤ x ≤ 0.3 and shifts dramatically to lower angle when x ≥ 0.35. Similar behavior can be found for the (1 1 1) peak while opposite trend can be observed for the (2 0 0) peak, which initially shifts to low angle when x ≤ 0.3 and then to higher angle when x > 0.3/ In order to visualize the changes of the $V_{1-x}Ru_xO_2$ unit cell, lattice parameters obtained from Rietveld fitting are plotted in Figure 2(a) & 2(b). The results are in good agreement with previously reported lattice parameters.[43] An obvious trend of increasing length of $a$, which is directly relevant to the inter-chain distances, can be seen prior to x = 0.3 followed by a continuous decrease, i.e., an increase of 1.4% compared with the length of $a$ of $V_{0.97}Ru_{0.03}O_2$ is seen. In the meantime, the length of $c$, which describes the intra-chain V-V distances, shows different behavior, initially decreasing for x ≤ 0.15, i.e., a drop of 0.1%, and increases when x ≥0.20, i.e., an increase of 9.1% when compared to the length of $c$ for $V_{0.97}Ru_{0.03}O_2$.

For comparison purposes, to determine whether this behavior is seen for other dopants of $VO_2$, the cell parameters for a different chemical origin of the electron doping, for $V_{1-x}Mo_xO_2$, were extracted from ref. 21 and plotted in Figure 2(a). This allows one to determine whether it is the valence electron count (VEC) per atom that primarily impacts the lattice parameters. Unlike the case for $V_{1-x}Ru_xO_2$, $a$ in $V_{1-x}Mo_xO_2$ increases monotonically with increasing Mo content while the $c$ first increases and then drops, which is opposite to what is seen for $V_{1-x}Ru_xO_2$. The trends of $c/a$ and volume of the unit cell for both $V_{1-x}Ru_xO_2$ and $V_{1-x}Mo_xO_2$ are presented in Figure 2(c). When the VEC per atom is smaller than 5.967 e⁻/atom (x < 30% in $V_{1-x}Ru_xO_2$), $c/a$ of $V_{1-x}Ru_xO_2$ decreases with increasing x, the same as $V_{1-x}Mo_xO_2$. Beyond this point, the $c/a$ ratio of $V_{1-x}Ru_xO_2$ increases with higher concentration of Ru. The inset of Figure 2(c) illustrates the volume of the unit cell with respect to VEC/atom. Consistent with the larger ionic radius of Mo and Ru than V, the volumes of both series of materials increase as the amount of dopant increases. It is clear from the Mo-Ru doping comparison that the VEC/atom is not the factor that primarily affects the lattice dimensions of $V_{1-x}M_xO_2$. Therefore, we compare the influence of doping value x by comparing the lattice parameters, cell volume and c/a ratio of $V_{1-x}M_xO_2$, as illustrated in Figure 2(b) & 2(d). Similar to what is observed for the VEC/atom case, "x" (the concentration of dopant) also does not play an essential role in determining lattice dimensions in the doped $VO_2$ system.

**3.2. Magnetic Properties:** The general magnetic properties for $V_{1-x}Ru_xO_2$ are shown in Figure 3. The temperature-dependence of the magnetic susceptibility is presented in Figure 3(a). Sharp transitions reflecting the MIT can be observed above 250 K for x < 0.10, as shown in the main panel of Figure 3(a). The transition temperatures are visualized by the peaks of the first derivatives of χT vs T curves shown in the inset. The MIT for x = 0.01, 0.02, 0.03, 0.04, 0.05, 0.06, 0.08 and 0.10 are seen at ~327 K, ~312 K, ~299 K, ~289 K, ~280 K, ~270 K, ~255 K and ~247 K, which is shown in Figure 4(a). For x > 0.1 no peak can be seen in d(χT)/dT vs T since the magnetic susceptibility curves, presented in the top left corner, become smoothly varying with temperature. Interestingly, for x = 0.01, the magnetic susceptibility is smaller than that of any other sample immediately below the MIT temperature. This might be due to the fact that, similar to the case for undoped $VO_2$, the formation of spin singlet below the MIT temperature weakens the magnetic response to an external magnetic field. In Figure 3(b), the temperature-dependence of the inverse magnetic susceptibility ($\chi^{-1}$) below 30 K indicates Curie-Weiss-like local moment behavior and no indication of long-range magnetic ordering can be found. Clear Curie-Weiss behavior can be seen for undoped $VO_2$ and Mo-doped $VO_2$ at low temperature, as shown in ref. 21. However, by doping Ru atoms onto V site, the $\chi^{-1}$ vs T curves are bent away from the linear fitting for the temperature range from 20 K to 30 K, as indicated by the solid lines. The linear fitting is based on the Curie Weiss law: $1/\chi = T/C - \theta_{CW}/C$, where χ is magnetic susceptibility, $\theta_{CW}$ is Curie-Weiss temperature and C is a constant from which the effective moment ($\mu_{eff} = \sqrt{8C}$ $\mu_B$) is derived. Thus, when x increases from 0.01 to 0.10, $\mu_{eff}$ of $V_{1-x}Ru_xO_2$ increases, and when 0.10 < x ≤ 0.30, $\mu_{eff}$ drops monotonically then to rise again when x > 0.30. In addition, $\theta_{CW}$ is negative, ranging from -0.52 (7) K (at x = 0.01) to -353 (9) K (for x = 0.09), which reveals that antiferromagnetic interactions are dominant in the $V_{1-x}Ru_xO_2$ solid solution in the corresponding temperature range.

The field-dependent magnetization (MH) curves from 0 to 9 T for $V_{1-x}Ru_xO_2$ at 1.8 K are shown in Figure 3(c) & (d). Although the solid solution shows no evidence for long-range magnetic ordering, all the curves are found to bend towards the H axis when the applied magnetic field is large enough, i.e., $\mu_0 H > 2$ T, and exhibit a small unsaturated magnetic moment at 9 T, which is summarized in Figure 4(b) The MH curves clearly demonstrate that the magnetic moments at 9 T increase with x when $x \leq 0.06$ and decrease with x when $x > 0.06$. The linear behavior of MH curves for $x = .70$ and $0.90$ is closer to what is expected for normal paramagnetic materials.

**3.3. Resistivity:** The temperature dependent resistivity from 1.8 K to 300 K in the absence of an applied magnetic field was measured on polycrystalline pellets. Figure 5 presents the temperature-dependence of normalized resistivity ($\rho(T)/\rho(350)$). As can be seen in the main panel, a sharp increase in resistivity can be found for materials with smaller x values ($x < 0.15$), due to the MIT. The transition temperature is gradually suppressed to lower temperature with larger x while the MIT cannot be seen in the polycrystalline samples when x reaches 0.15. Thus, the sharp MIT disappears in temperature-dependent resistivity measurements at higher x, but the material remains poorly conducting. Although Ru atoms, which have three more valence electrons than V, are introduced into the system to provide more conducting electrons, the materials persist in their insulating behavior until a surprisingly large doping value. When x increases to 0.50, as shown in the inset of Figure 5, the material exhibits clear insulating behavior. However, when $x = 60\%$, such insulating behavior is dramatically suppressed, and a metallic resistivity curve is eventually obtained for $x = 80\%$. Thus, low-temperature resistivity measurements down to 0.1 K for $x = 80\%$ and 90% were carried out; no evidence for bulk superconductivity was observed.

**3.4. Heat Capacity:** To better probe the electronic behavior of $V_{1-x}Ru_xO_2$, low-temperature heat capacity measurements from 2 K to 30 K under no applied magnetic field were performed. The results are presented in Figure 6. Both axes in the inset are shown on a logarithmic scale for clarity of the data below 10 K while linear scale is used in the main panel. No obvious heat capacity jump corresponding to any kind of phase transition can be seen in the full measured temperature range, which is consistent with the magnetic susceptibility and resistivity measurements. Generally, the heat capacity $C_p$ of materials at low temperature can be represented by $C_p = \gamma T + \beta T^3$ where $\gamma$ and $\beta$ reflect the electronic (Sommerfeld coefficient) and phononic contributions, respectively. Thus, a linear fitting for $C_p/T$ *vs* $T^2$ curve can be applied to estimate the $\gamma$ and $\beta$ in $V_{1-x}Ru_xO_2$, as shown by solid lines in Figure 6. The intercept of the fitting line with y axis yields the value of $\gamma$. It can be clearly observed that when $x \leq 10\%$, $\gamma$ decreases with increasing x from 7.1 (2) mJ/(mol K$^2$) ($x = 1\%$) to 0.4 (2) mJ/(mol K$^2$) ($x = 10\%$). Furthermore, as x increases from 10% to 90%, $\gamma$ increases slowly, to 7.5 (1) mJ/(mol K$^2$) (x = 90%). For insulators, γ is usually close to zero due to the existence of the bandgap, such as the small γ (~1.0 mJ/(mol K$^2$)) of undoped VO$_2$[21]. Interestingly, although the current materials exhibit insulating behavior when x < 80%, the Sommerfeld coefficient is still growing as x increases. Such behavior indicates that by introducing more valence electrons into VO$_2$, the strength of electronic fluctuations varies a lot. Moreover, the low-temperature upturns observed in the heat capacity data can be attributed to Schottky anomalies arising from the magnetic moments' fluctuations, similar to what is seen for Mo-doped VO$_2$.[21]

**3.5. Discussion:** To investigate whether the behavior for V$_{1-x}$Ru$_x$O$_2$ is similar to that for another chemically-doped VO$_2$ systems, V$_{1-x}$Mo$_x$O$_2$, a comparison of γ and μ$_{eff}$ for both systems is seen in Figure 7(a). (The data for the V$_{1-x}$Mo$_x$O$_2$ system are extracted from ref. 21.) The data are shown plotted both as valence electron count (VEC) per atom and dopant concentration x per atom. When the VEC/atom hits ~5.77, V$_{1-x}$Ru$_x$O$_2$ presents a broad minimum for γ, while V$_{1-x}$Mo$_x$O$_2$ has an opposite behavior, i.e., a peak. Moreover, when the doping value changes, a broad minimum of γ appears near x = 0.10 for V$_{1-x}$Ru$_x$O$_2$, increasing and reaching a peak near x = 0.23. For μ$_{eff}$, a broad maximum emerges at VEC/atom ~5.75 in V$_{1-x}$Ru$_x$O$_2$ while for V$_{1-x}$Mo$_x$O$_2$, μ$_{eff}$ first drops to a minimum at VEC/atom ~5.7 and then reaches a peak at VEC/atom ~5.72 followed by a valley at VEC/atom ~5.75, which is opposite to what is seen for the Ru counterpart. Furthermore, when the concentration of dopant changes, when x = 0.10, V$_{1-x}$Ru$_x$O$_2$ exhibits a broad maximum while V$_{1-x}$Mo$_x$O$_2$ shows a minimum.

Different trends between the two systems for both γ and μ$_{eff}$ are observed, which indicates that neither the VEC/atom nor the doping value x are the significant factors that decide the electronic properties of electron-doped VO$_2$. Therefore, a specific picture is needed to describe how Ru doping can affect the physical properties in V$_{1-x}$Ru$_x$O$_2$. A previous report claims that VO$_2$ has an electronic configuration where t$_{2g}$ orbitals are formed by one fully filled orbital and two half-filled orbitals instead of two fully filled orbitals which cannot provide unpaired electrons.[28] Thus, with that information in mind, by integrating magnetic and electronic transport behaviors, it can be speculated that when x ≤ 6%, the magnetic moment at 9 T increases due to the fact that Ru brings more electrons into the system and breaks the spin singlet in VO$_2$ that can suppress the MIT to lower temperature and produce a local magnetic moment. When the doping value increases further, to ~10%, the Peierls distortion takes over and starts to become dominant and, thus, makes a small amount of spin singlet states. This assumption can be evidenced by the Wilson ratio ($R_W$), defined as $R_W = \frac{4\pi^2 \chi_0}{3\gamma}$, where $\chi_0$ is related to the core diamagnetism and temperature independent paramagnetic contributions such as Pauli paramagnetism and is obtained by fitting temperature-dependent magnetic susceptibility by using modified Curie-Weiss law $\chi = \chi_0 + C/(T-\theta_{CW})$, shown in Figure 7(b).[45–47] $R_W$ quantifies the spin fluctuations that enhance the magnetic susceptibility. In $V_{1-x}Ru_xO_2$, $R_W$ starts at ~7.3 when x = 1% and dramatically increases to ~1460 when x = 10%, dropping gradually to ~300 beyond this point. This means that spin fluctuations become stronger when the Ru doping is less than 10%, suggesting that the ferromagnetism originates from Ru 4$d$ electrons; considering that $SrRuO_3$ with $Ru^{4+}$ is ferromagnetic.[45,47, 48] After 10%, the spins start to re-form the spin singlet state and the magnetization and $R_W$ both drop as a consequence of weaker spin fluctuations.[45,47] This process can lead to both insulating behavior and decreasing of the magnetic moment, continuing to x ~ 80%. When x ≥ 80%, the metallic nature of $RuO_2$ eventually prevails, resulting in poor metal behavior for heavily doped $V_{1-x}Ru_xO_2$. Thus, we propose a phase diagram for Ru-doped $VO_2$ shown as Figure 8. The circles stand for the MIT temperature observed in temperature-dependent magnetic susceptibility. The white area means that the MIT cannot be observed any more. With heavier doping of Ru, $V_{1-x}Ru_xO_2$ changes from a Curie-Weiss insulator to a singlet insulator and eventually turns into a metal above x ~80%. In the meantime, superconductivity was not observed above 0.1 K for metallic samples.

## *4. Conclusion*

In this paper, we present the synthesis of a series of materials $V_{1-x}Ru_xO_2$ (1% ≤ x ≤ 90%) using the high-temperature solid-state method. The magnetic properties, electronic transport properties and heat capacity of selective compositions are investigated. The physical properties of $V_{1-x}Ru_xO_2$ are compared with previously reported $V_{1-x}Mo_xO_2$ and obvious differences can be observed between the two systems as a function of either the valence electron count (VEC) per atom or the doping value x, which implies neither of them can be the most significant factor in deciding the electronic/magnetic behavior in electron-doped $VO_2$. By integrating the magnetic and electronic behaviors and the Wilson ratio, we propose a possible reason for the behavior of $V_{1-x}Ru_xO_2$, but deeper interpretation of the phenomena observed requires further study, such as through the determination of the local structure and further theoretical study of the relationship between itinerant and localized behavior in the rutile structure.

## *Acknowledgements*

This research was supported by the Gordon and Betty Moore Foundation, EPIQS initiative, grant GBMF-9066.

## *Supplementary Information*

Supplementary information for this article can be found online at xxxxxxx.

**Figure 1. (a)** The crystal structure of $V_{1-x}Ru_xO_2$ where blue and cyan balls represent V atoms and pink and red balls stand for O atoms. The bottom figure shows the linear chain feature of V in $VO_2$. **(b) (Main panel)** Powder X-ray diffraction patterns of $V_{1-x}Ru_xO_2$ at 293 K; **(Inset) (Left)** The trend of (0 0 2) peak position marked by black arrows; **(Inset) (Right)** The trends of (2 0 0) and (1 1 1) peak positions.

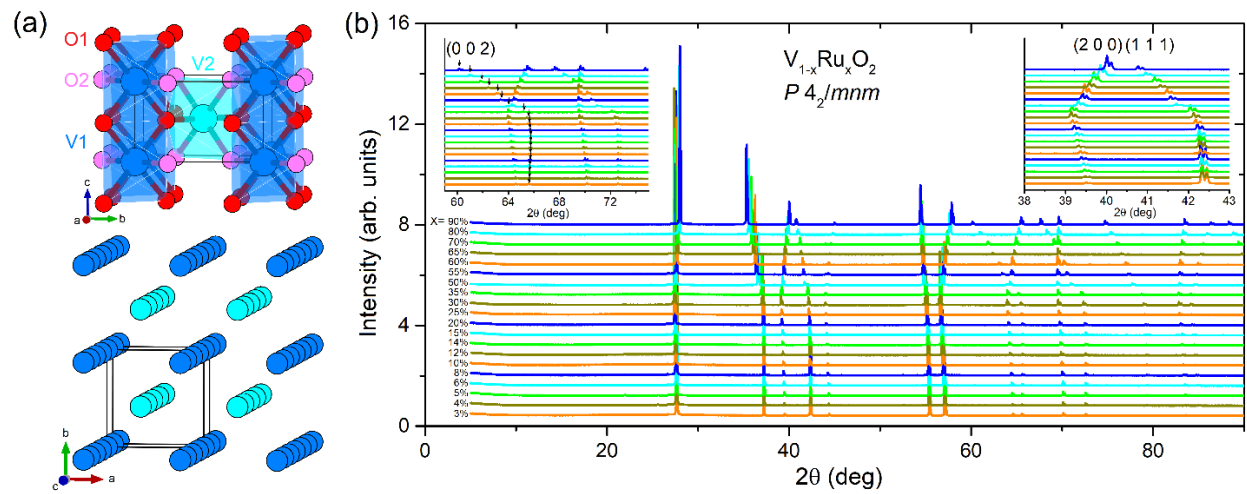

**Figure 2 (Main panel)** The lattice dimensions of $V_{1-x}M_xO_2$ (M = Ru or Mo) with respect to **(a)** valence electron count (VEC) per atom & **(b)** x in $V_{1-x}M_xO_2$. **(Inset)** Zoom-in of the dimension of *c* in the rutile unit cell for $V_{1-x}M_xO_2$. **(Main panel)** *c/a* of $V_{1-x}M_xO_2$ with respect to **(c)** VEC/atom & **(d)** x in $V_{1-x}M_xO_2$; **(Inset)** Cell volume of $V_{1-x}M_xO_2$ with respect to **(c)** VEC/atom & **(d)** x in $V_{1-x}M_xO_2$.

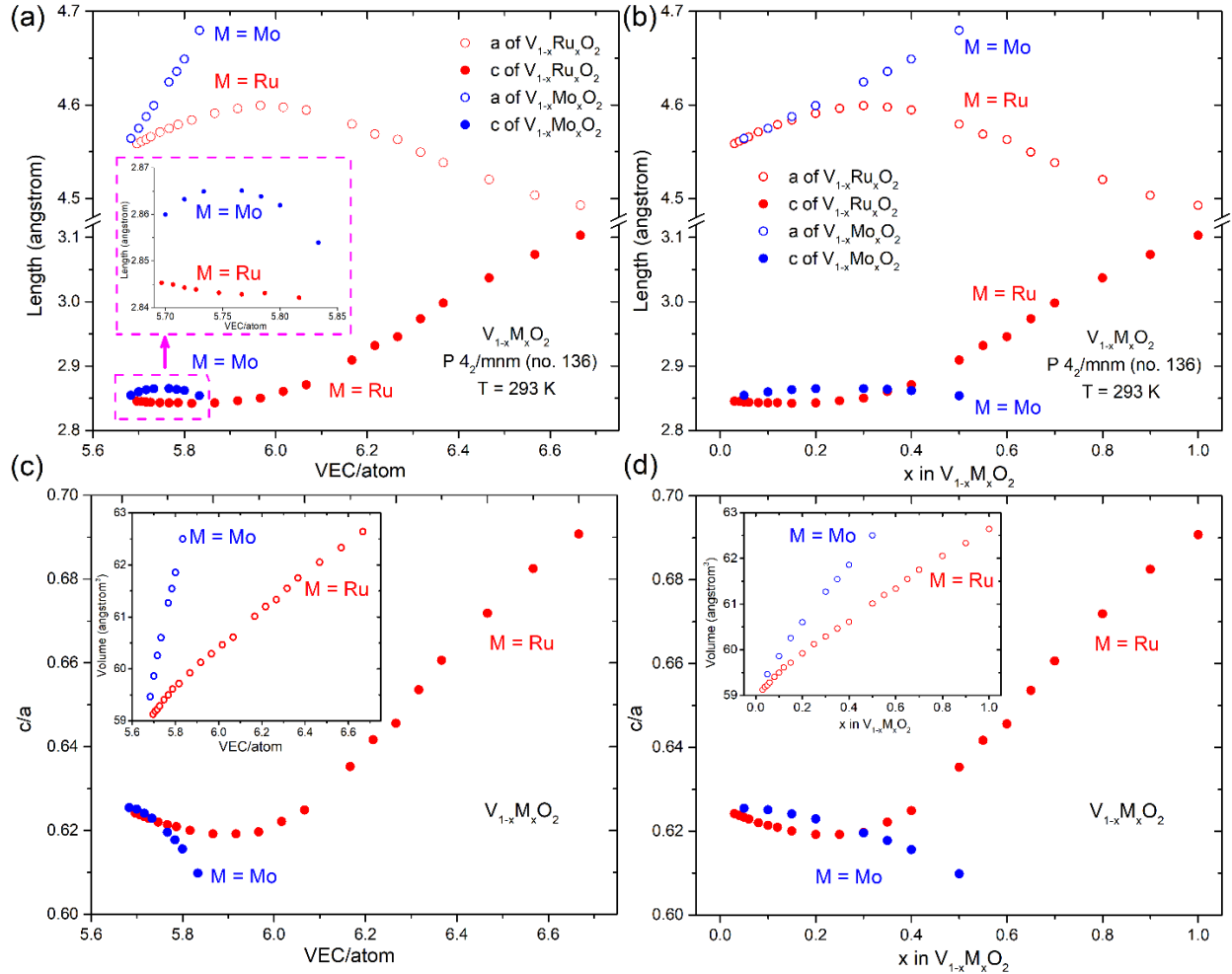

**Figure 3. (a)** The temperature-dependence of the magnetic susceptibility of $V_{1-x}Ru_xO_2$ for **(Main panel)** $x \leq 10\%$ and **(Inset) (Left)** $x > 10\%$; **(Inset) (Right)** $d(\chi T)/dT$ *vs* T curves. **(b)** Temperature-dependence of inverse magnetic susceptibility for **(Main panel)** $x \leq 10\%$ and **(Inset)** $x > 10\%$, the solid lines are linear fittings to the Curie-Weiss law. The field-dependence of the magnetization for $V_{1-x}Ru_xO_2$ with **(c)** $x \leq 6\%$ & **(d)** $x \geq 6\%$.

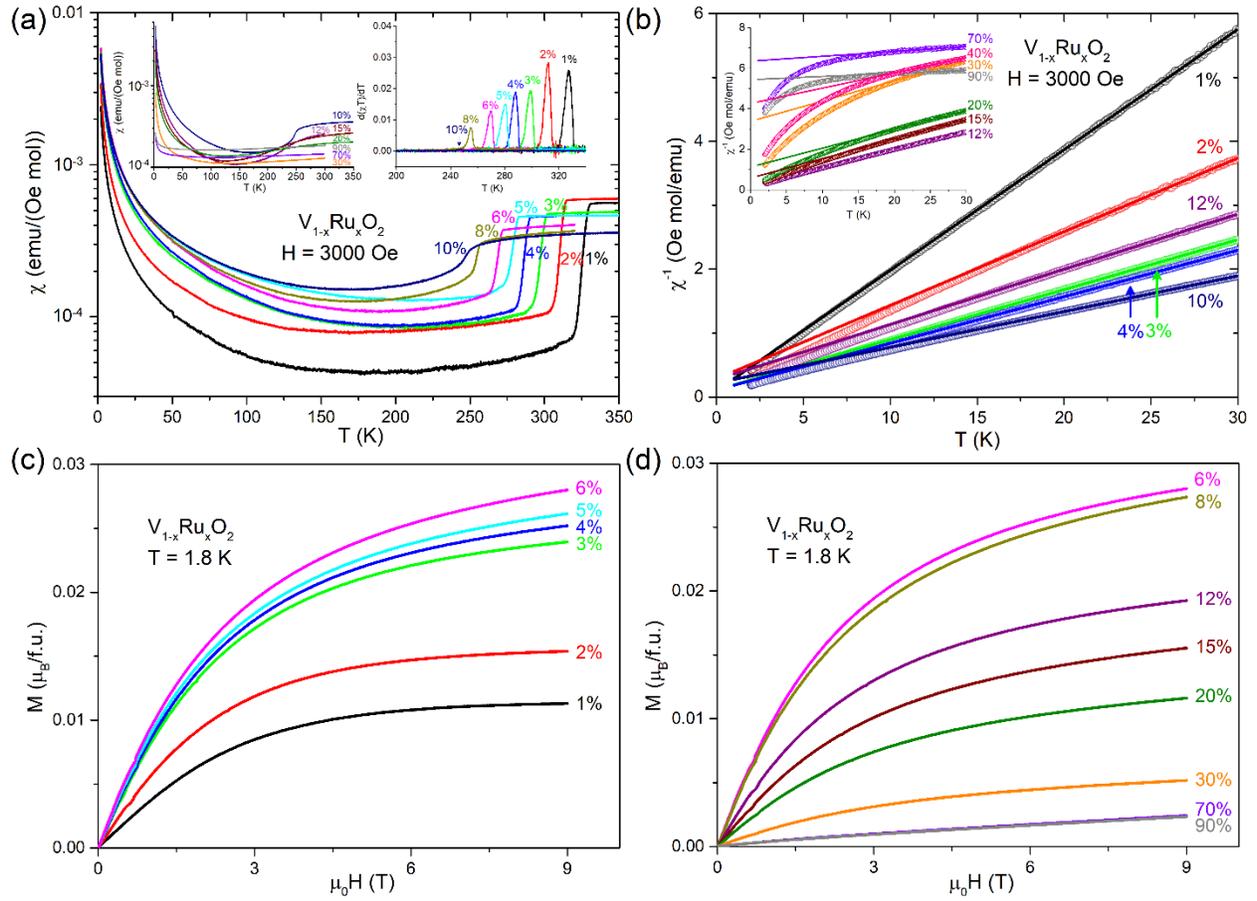

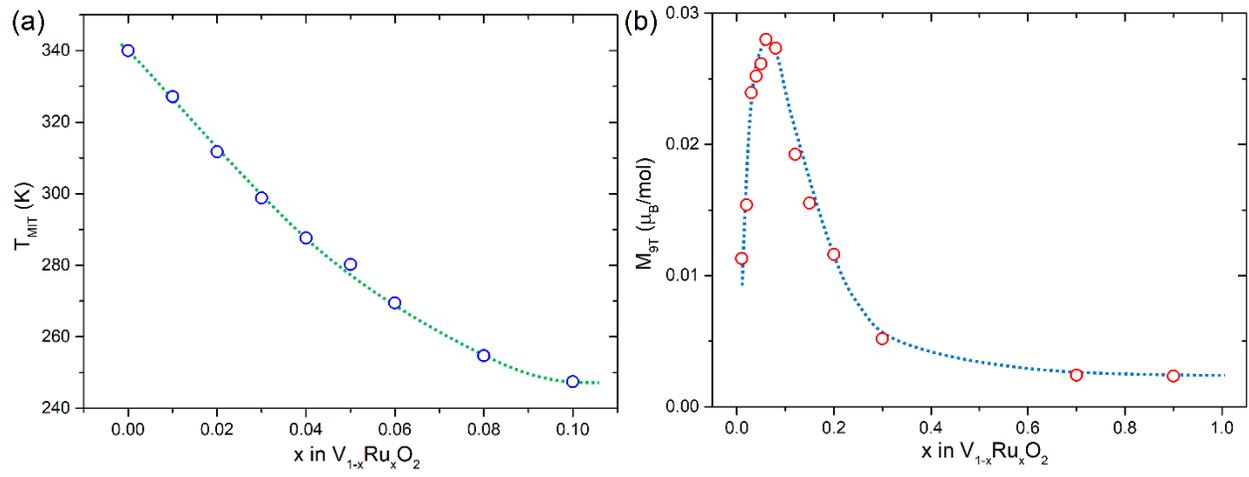

**Figure 4. (a)** The trend of the MIT temperature ($T_{MIT}$) with respect to x in $V_{1-x}Ru_xO_2$. **(b)** Molar magnetization at 9 T and 1.8 K with respect to x in $V_{1-x}Ru_xO_2$. The dotted lines in both figures are guides to the eye.

**Figure 5.** The temperature-dependence of the normalized resistivity of $V_{1-x}Ru_xO_2$ for **(Main panel)** x ≤ 50% and **(Inset)** x ≥ 50%.

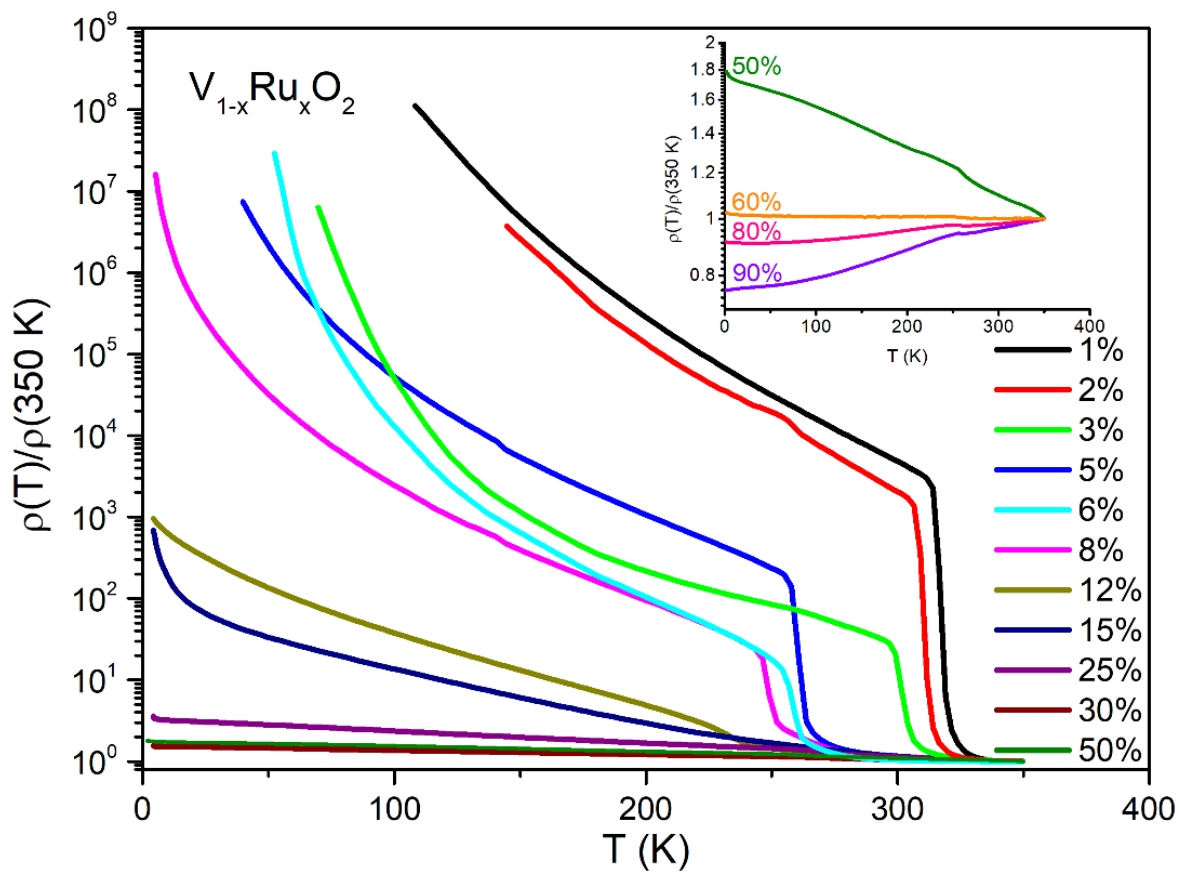

**Figure 6. (Main panel)** $C_p/T$ *vs* $T^2$ curves with a linear scale on both axes. **(Inset)** $C_p/T$ *vs* $T^2$ curves with logarithmic scale on both axes. The solid lines are for linear fitting of the higher-temperature data from 400 $K^2$ to 900 $K^2$ (20-30 K).

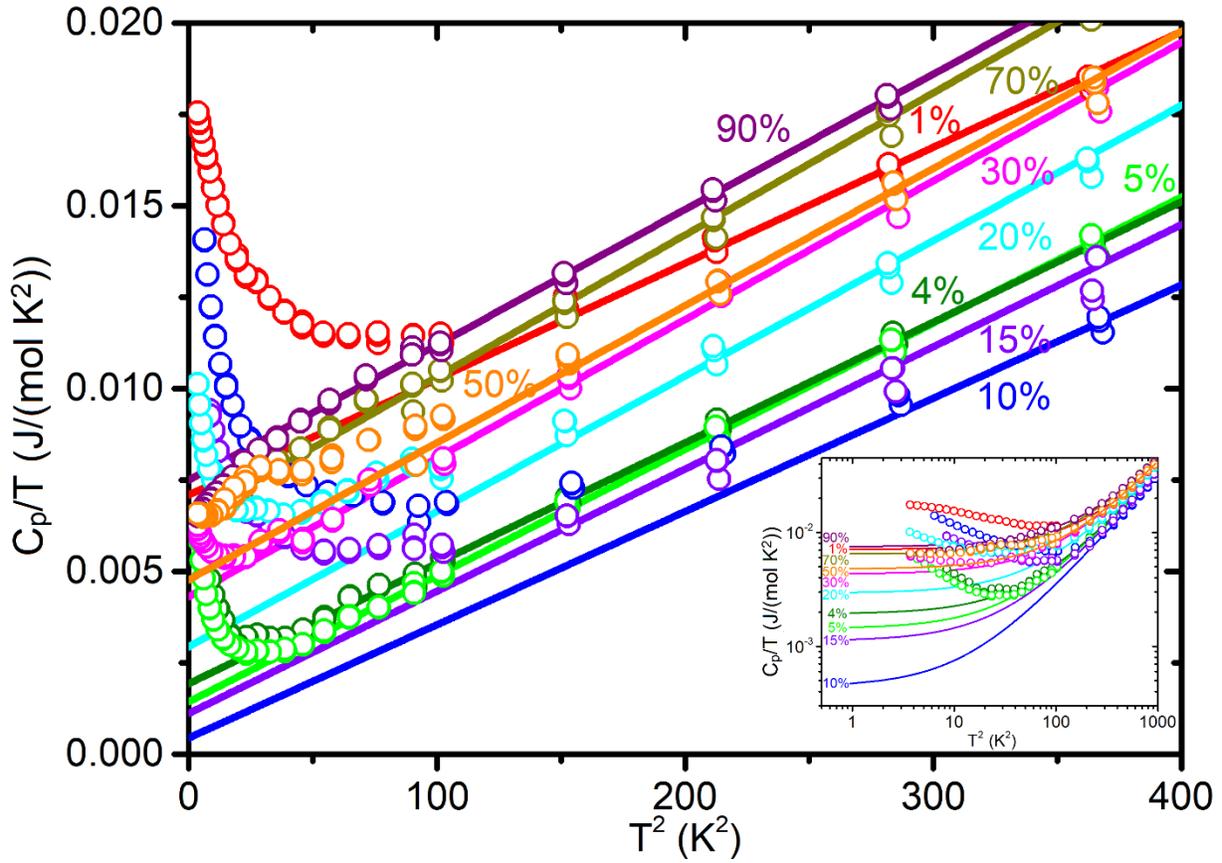

**Figure 7. (a) (Upper panel)** Sommerfeld coefficient g and **(Lower panel)** effective moment $\mu_{eff}$ for $V_{1-x}M_xO_2$ (M = Mo/Ru) with respect to **(Main panel)** VEC/atom and **(Inset)** x in $V_{1-x}M_xO_2$. The black/magenta solid & dashed lines are guides to the eye. **(b)** The Wilson ratio of $V_{1-x}Ru_xO_2$. The black dotted line is a guide to the eye.

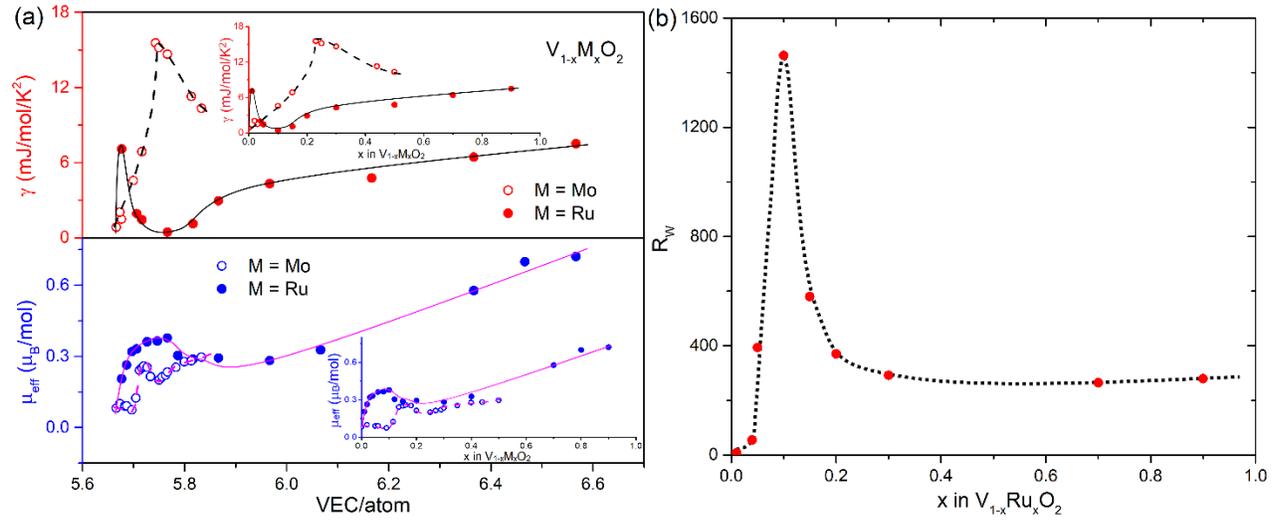

**Figure 8.** Proposed electronic phase diagram for $V_{1-x}Ru_xO_2$.

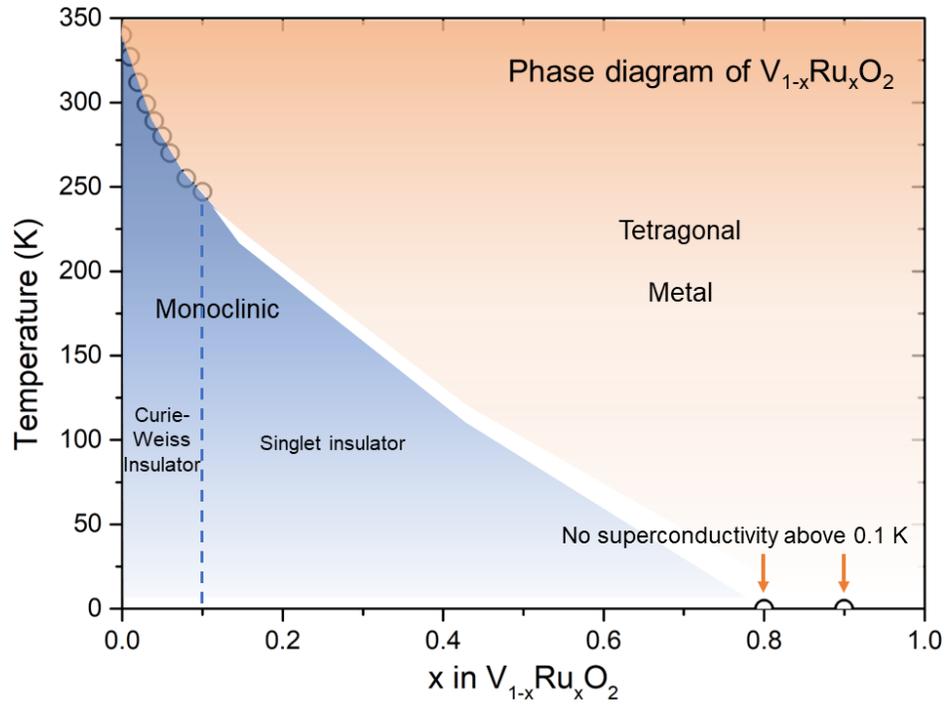

Supporting Information for

# Metal-Insulator Transition and Anomalous Lattice Parameters Changes in Ru-doped $VO_2$

*Xin Gui and Robert J. Cava*[*]

Department of Chemistry, Princeton University, Princeton NJ 08540, USA

*Corresponding author: R. J. Cava: rcava@princeton.edu

**Table of Contents**

**Figure S1.** Powder XRD patterns for $V_{0.99}Ru_{0.01}O_2$ and $V_{0.98}Ru_{0.02}O_2$ with Le Bail fitting.

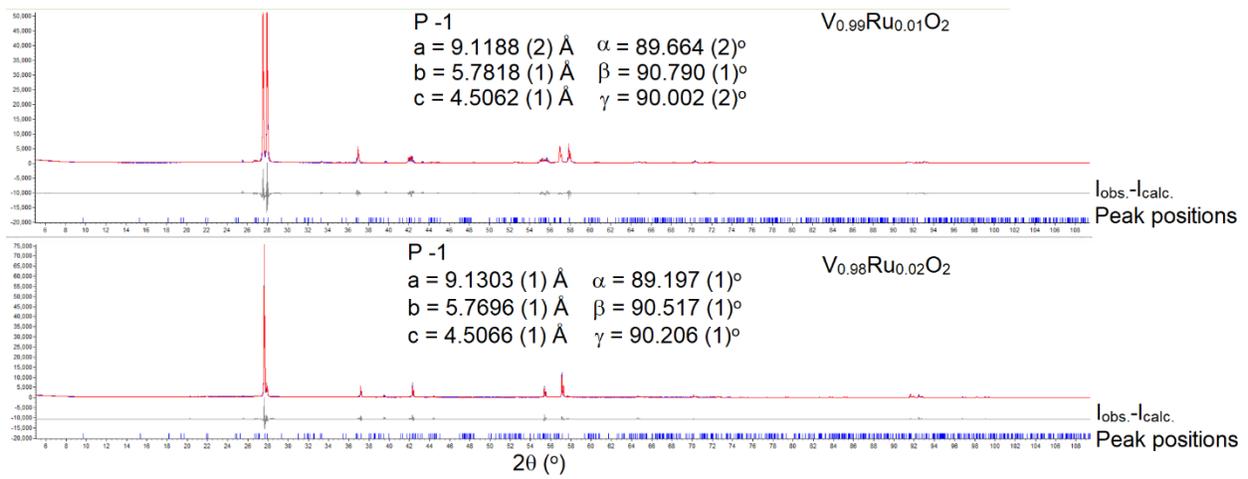

**Figure S2.** Powder X-ray diffraction pattern of $V_{1-x}Ru_xO_2$ (x ≥ 3%) with Rietveld fitting where black lines are observed patterns and red lines stand for calculated patterns.

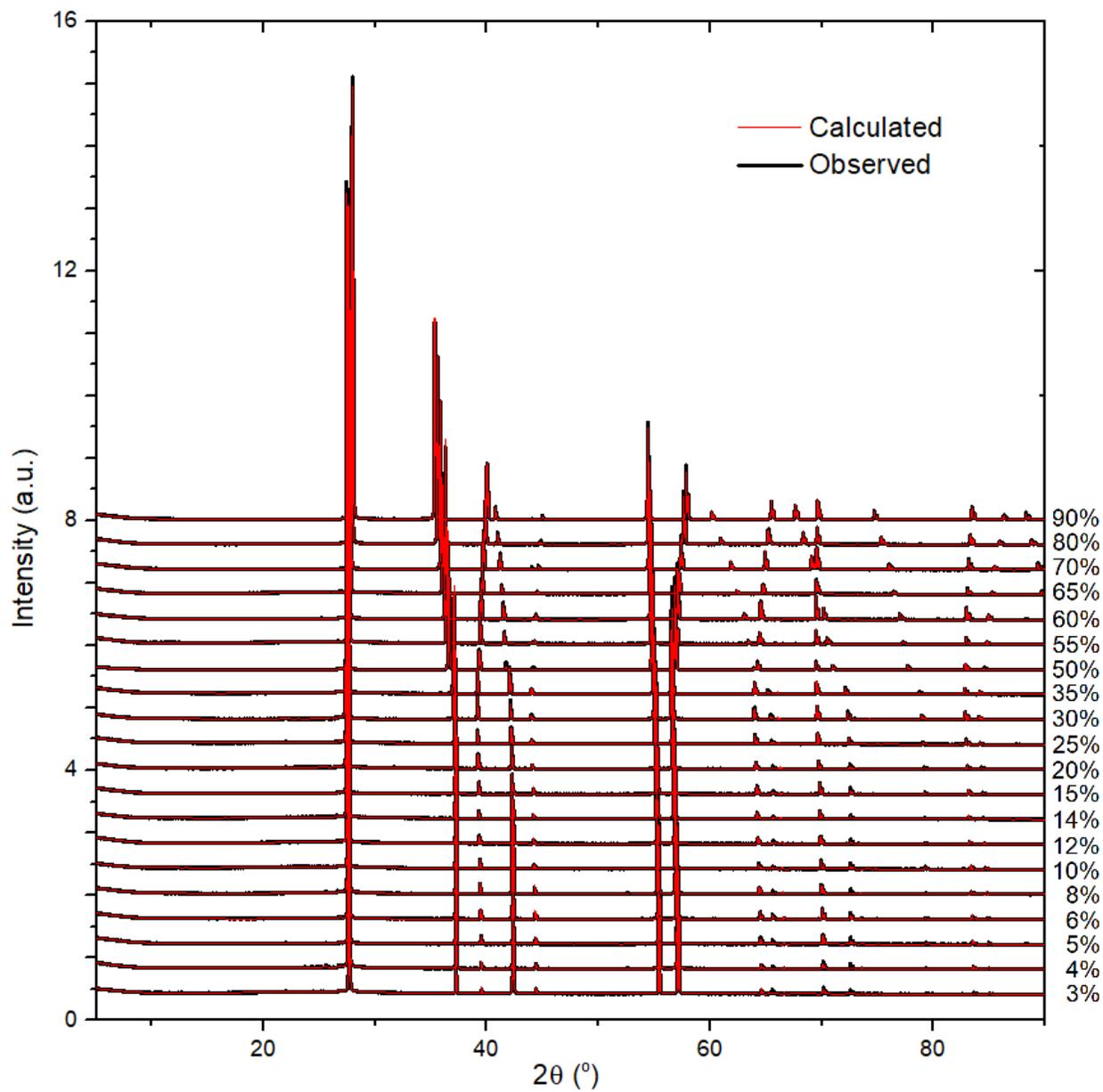